\documentclass[%
 reprint,
superscriptaddress,
nofootinbib,
 amsmath,amssymb,
 aps,
]{revtex4-1}
\usepackage{mhchem}
\usepackage[utf8]{inputenc}
\usepackage[T1]{fontenc}
\usepackage{amsmath}
\usepackage[normalem]{ulem}
\usepackage{dcolumn}
\usepackage{bm}
\usepackage{makecell}
\usepackage{multirow}
\usepackage{xcolor}
\usepackage{graphicx}
\usepackage{dcolumn}
\usepackage{bm}
\usepackage{gensymb}
\usepackage{siunitx}
\usepackage{booktabs}
\usepackage{caption}
\usepackage{subcaption}
\usepackage{setspace}
\usepackage{float}
\usepackage{caption}
\usepackage{array}
\usepackage{lipsum}
\usepackage{color}
\usepackage{xcolor}

\usepackage{varioref}       
\usepackage{hyperref}      
\usepackage{cleveref}     

\crefname{table}{table}{tables}
\Crefname{table}{Table}{Tables}
\crefname{figure}{Fig.}{Figs.}
\Crefname{figure}{Figure}{Figures}

\newcolumntype{P}[1]{>{\centering\arraybackslash}p{#1}}

\newcommand{\entireD}{222,960}

\newcommand{\iterI}{700}
\newcommand{\screenI}{221,990}

\newcommand{\iterII}{1970}

\newcommand{\iterIII}{3593}
\newcommand{\screenIII}{213,200}

\newcommand{\entiresource}{500}

\newcommand{\sourceABXD}{500}
\newcommand{\sourceABXDEg}{487}

\newcommand{\NtrainEg}{766}
\newcommand{\NtestEg}{180}
\newcommand{\entireDEg}{946}

\newcommand{\known}{8}

\newcommand{\subsetD}{48}

\newcommand{\Eg}{$E_{g}$}
\newcommand{\Ef}{$\Delta H_{f}$}
\newcommand{\EgHSE}{$E_{g}^{HSE}$}
\newcommand{\EgPBE}{$E_{g}^{PBE}$}
\newcommand{\EgHSEML}{$E_{g}^{ML, HSE}$}
\newcommand{\EgPBEML}{$E_{g}^{ML, PBE}$}
\newcommand{\EfPBE}{$\Delta H_{f}^{PBE}$}
\newcommand{\EfPBEML}{$\Delta H_{f}^{ML, PBE}$}
\newcommand{\EdPBE}{$\Delta H_{d}^{PBE}$}
\newcommand{\EdPBEML}{$\Delta H_{d}^{ML, PBE}$}

\newcommand{\AL}{ALIGNN}
\newcommand{\Sch}{SchNet}
\newcommand{\PN}{PaiNN}

\newcommand{\ALerrEf}{38 meV/atom}
\newcommand{\ALerrEg}{0.18 eV}

\begin{document}

\title {Leveraging Domain Adaptation for Accurate Machine Learning Predictions of New Halide Perovskites}

\author{Dipannoy Das Gupta}
\affiliation{Department of Chemistry and Biochemistry, University of South Carolina, Columbia, SC 29208, United States}
\affiliation{Department of Computer Science and Engineering, University of South Carolina, Columbia, SC 29208, United States}

\author{Zachary J. L. Bare}
\affiliation{Department of Chemistry and Biochemistry, University of South Carolina, Columbia, SC 29208, United States}

\author{Suxuen Yew}
\affiliation{Department of Chemical and Biological Engineering, University of Colorado, Boulder, CO 80303, United States}

\author{Santosh Adhikari}
\affiliation{Department of Chemistry and Biochemistry, University of South Carolina, Columbia, SC 29208, United States}

\author{Brian DeCost}
\affiliation{National Institute of Standards and Technology, Gaithersburg, MD 20899, United States}

\author{Qi Zhang}
\affiliation{Department of Computer Science and Engineering, University of South Carolina, Columbia, SC 29208, United States}

\author{Charles Musgrave}
\affiliation{Department of Chemical and Biological Engineering, University of Colorado, Boulder, CO 80303, United States}

\author{Christopher Sutton}
\affiliation{Department of Chemistry and Biochemistry, University of South Carolina, Columbia, SC 29208, United States}
\email{cs113@mailbox.sc.edu}

\begin{abstract}
We combine graph neural networks (GNN) with an inexpensive and reliable structure generation approach based on the bond-valence method (BVM) to train accurate machine learning models for screening \entireD{} halide perovskites using statistical estimates of the DFT/PBE formation energy (\EfPBE{}), and the PBE (\EgPBE{}) and HSE (\EgHSE{}) band gaps. The GNNs were fined tuned using domain adaptation (DA) from a source model, which yields a factor of 1.8 times improvement in \EfPBE{} and 1.2 - 1.35 times improvement in \EgHSE{} compared to direct training (i.e., without DA). As far as we are aware, this is the first demonstration of DA for GNNs in materials science. Moreover, after three active learning iterations -- resulting in a total of \iterIII{} training samples for \Ef{} --  the best ML model produced mean absolute errors (MAE) of \ALerrEf{} relative to \EfPBE{} for totally unseen materials. The ML model for predicting \EgHSE{} achieved an MAE of \ALerrEg{}. Using these two ML models, \subsetD{} compounds were identified out of \entireD{} candidates as both stable and that have an HSE band gap that is relevant for photovoltaic applications. For this subset, only \known{} have been reported to date, indicating that 40 compounds remain unexplored to the best of our knowledge and therefore offer opportunities for potential experimental examination. 
\end{abstract}

\maketitle

\section{Introduction}

Halide perovskites are a broadly used class of materials for a variety of applications, including light-emitting diodes (LEDs) and photocatalysis that are particularly promising as light-adsorbing materials for photovoltaics (PV).\cite{yin2015halide,ansari2018frontiers,huang2020solar, jacobsson2022open} One reason for the wide applicability of perovskites is their extensive chemical space,\cite{bartel2020inorganic} which enables modulation of their optical, electronic, and thermodynamic properties by changing their compositions. Although the enormous chemical space opens up opportunities for fine-tuning the properties of potential materials, it simultaneously poses a significant challenge for effectively screening materials.

Recent improvements in computing power allow for the use of quantum mechanical calculations, e.g., density functional theory (DFT), for high-throughput screening of materials. Specifically for screening perovskites, Kar et al. and Schmidt et al. have examined \ce{ABX3} compounds.\cite{kar2020computational,yang2022high,schmidt2017predicting} However, AA'BB'X$_6$ perovskites are less explored because their unit cells are at least twice as large as those of \ce{ABX3} perovskites, and thus, approximately 8 times more computationally expensive when assuming cubic scaling (N$^3$) of DFT. Alternatively, machine learning (ML) can be used to accurately predict DFT properties of atomistic systems at orders of magnitude lower computational cost, thereby enabling the rapid evaluation of materials over a large chemical space for promising candidates.

The accuracy of an ML model is significantly influenced by the selection of input features or representation. Because ML-based high-throughput screening tasks typically cannot include features from the DFT-optimized structure, tabulated atomic properties (such as radii) have been utilized as the inputs in the ML models used to screen materials generally,\cite{goodall2020predicting, jha2018elemnet,ward2018matminer} and specifically for halide perovskites.\cite{yang2022high, tao2021machine, im2019identifying, gao2021screening, wu2019global, diao2022high} Some of us have previously reported a descriptor,\cite{bartel2019new} $\tau$ that allows for the classification of whether a compound will form a stable perovskite based only on its composition. Because $\tau$ and other composition-only models use only tabulated atomic information, it estimates stability very efficiently. 

However, composition-based ML models have limitations in describing properties (e.g., the stability and band gaps), which can vary significantly depending on the atomic structure. This has been shown in the literature for perovskites specifically,\cite{zhao2020polymorphous,baldwin2023dynamic,prasanna2017band,varignon2019origin, bartel2020inorganic} where local symmetry breaking resulting from BX$_6$ octahedral tilting and distortions significantly affects material properties. Therefore, models that incorporate structural information directly are preferred in order to account for this complication. 

Numerous structure-based representations have been developed for materials, which aim to either quantify the local environment or catalog the bond distances and angles within the structure.\cite{bartok2013representing, rupp2012fast, montavon2012learning, huo2017unified, hansen2015machine} Alternatively, it is convenient to represent materials as a graph, where each atomic site is a node, and an edge corresponds to bonds between atoms (defined as two sites lying within a specified cutoff distance).\cite{schutt2018schnet, chen2019graph} Numerical features can then be generated from the graph, such as through the counting of different paths through the graph, which can then be combined with some regression algorithm.\cite{sutton2018nomad} Alternatively, the graph representations can be combined with message passing neural networks (i.e., GNNs), as first demonstrated -- to the best of our knowledge -- in Ref.~\cite{schutt2018schnet}. These are well-suited for featurizing structure-based properties by directly processing the positions and distances between atoms as inputs into a neural network. This capability significantly reduces the necessity for manual feature engineering and typically achieves high accuracy in predicting various properties of materials.\cite{xie2018crystal, chen2019graph, schutt2018schnet, schutt2021equivariant, choudhary2021atomistic, gasteiger2021gemnet, gasteiger2020directional, gasteiger2020fast, kong2022density, geiger2022e3nn}

However, a key challenge is generating training sets that include atomic structures without conducting a DFT calculation. This limitation is particularly significant in high-throughput screening for the discovery of currently \textit{unknown} materials not tabulated in existing datasets. An alternative approach to bypass the need for DFT-optimized structures as input in the ML model involves several recently reported GNN-based ML interatomic potentials (MLIPs). Notable examples include M3GNet\cite{chen_universal_2022} and CHGNet,\cite{deng2023chgnet} as well as the MACE architecture.\cite{batatia_mace_2023} All of these models were trained on snapshots of structures from DFT relaxations and thus, could be used to optimize the geometries of arbitrary new compounds.

An additional potential limitation of using GNNs is their requirement for a large number of samples to parameterize the model and achieve high accuracy. This presents a challenge when screening new materials, where the goal is often to explore a vast composition space with a minimal set of training samples. Hence, training a model with a scheme that allows for reusing information from other ML models trained to unrelated datasets would be highly beneficial.

To address the challenges posed by the availability of limited annotated data, an unsupervised domain adaptation (DA) approach has been adopted.\cite{goetz2022addressing} This previous work aimed to extract domain-invariant features from both the source and target domains. However, to the best of our knowledge, the DA approach has not yet been demonstrated for structure-based GNNs to predict material properties.

Here, we present a strategy that facilitates the application of GNNs to materials screening in three ways: 1) bypassing inputting the DFT-optimized structures by using an inexpensive and accurate proxy structure-generation scheme\cite{morelock2022bond}; 2) improving the learning efficiency of GNNs via DA; and 3) enhancing model generalization to a large unlabeled space using active learning. We apply all three strategies to screen \entireD{} AA'BB'$X_{6}$ halide perovskites based on three distinct targets that are relevant for the identification of 48 promising materials for PV applications: PBE-computed formation energies (\EfPBE{}) and band gaps (\EgPBE),\cite{pbe} and HSE-computed band gaps (\EgHSE).\cite{hse06} To perform a comparative analysis, three representative GNN architectures were trained in this work; \Sch{}, which has permutational, translational, and rotational invariance;\cite{schutt2018schnet} \PN{}, which includes translational and permutational invariances, as well as, rotational invariances and covariances;~\cite{schutt2021equivariant} and \AL{}, which incorporates both a line graph and a bond graph and has demonstrated strong performance across various benchmark datasets.\cite{choudhary2021atomistic}

\section{Results and Discussion}

\subsection{Dataset construction and preparation}\label{sssec:dcp}

\begin{figure}[htbp!]
    \captionsetup[subfigure]{labelformat=empty}
    \begin{subfigure}[b]{0.5\textwidth}
                \includegraphics[scale=0.15]{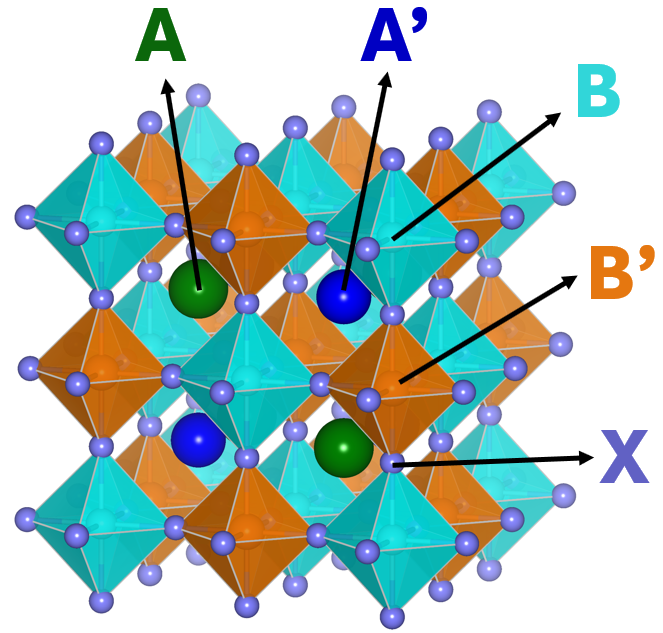}
                \caption{(a)}
    \end{subfigure}
        \begin{subfigure}[b]{0.5\textwidth}
                \includegraphics[scale=0.8]{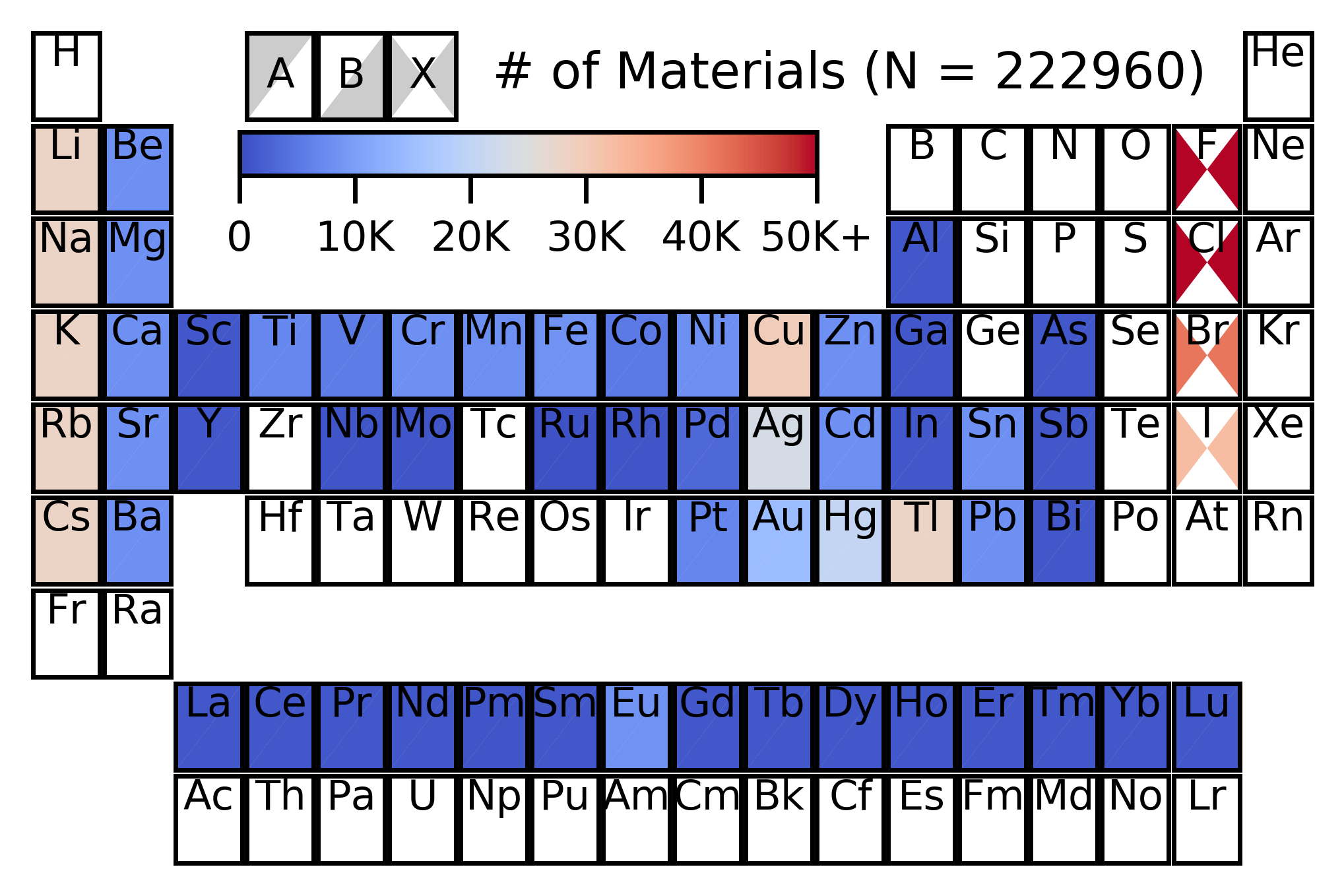}
                 \caption{(b)}
    \end{subfigure}

    \caption{(a) Cubic double perovskite (AA'BB'$X_{6}$) conventional unit cell structure with rock-salt site ordering. The larger A and A' sites are surrounded by corner-sharing B$X_{6}$ and B'$X_{6}$ octahedra. (b) Periodic table chart showing the A-site (upper left corner), B-site (lower right corner), and X-site (left and right triangles) frequencies (heatmap) of each element present in our dataset of \entireD{} halide AA'BB'X$_{6}$ compositions.}
       \label{fig:bvm_db}
\end{figure}

A total dataset of \entireD{} halide double perovskites was first generated by considering various combinations of 54 different elements occupying the A (A') and B (B') cation sites, and $X$ = F, Cl, Br, and I. To ensure that the combination of elements charge balance, the cation oxidation states were assigned using the algorithm reported by Bartel et al.\cite{bartel2019new} A heatmap illustrating the frequency of each element's occurrence on the A (A') and B (B') sites within the \entireD{} compositions is provided in \cref{fig:bvm_db}. For all of the double perovskites, a rock-salt cation ordering (see \cref{fig:bvm_db}A) within the cubic Fm-3m (space group \#225) was assumed. 

For all compositions, a cheap proxy structure was used as an input for the ML models, which was determined by minimizing the Global Instability Index (GII) following Ref. \cite{bare2023dataset}. These are referred to as BVM (bond valence method) structures throughout this work. The computational approach for generating the BVM structures is detailed in the Methods section.

The materials discovery workflow used in this work is as follows:

\begin{itemize}
    \item Active learning and DA were combined to efficiently train separate ML models using three different architectures for the prediction of \EfPBE{}, \EfPBEML{}, on a minimum set of training samples. 
    \item \EfPBEML{} was used to determine the decomposition energies (\EdPBE{}) for each of the \entireD{} AA'BB'$X_{6}$ compounds through the construction of a convex hull of all competing phases available in the OQMD database (see Methods section) to assess the stability relative to decomposition into other phases.
    \item A subset of \entireDEg{} AA'BB'X$_6$ compounds were selected based on specific criteria, and separate ML models were trained to predict the HSE06 band gap energies, \EgHSE{}. The training set for \EgHSEML{} was limited to a size that is practical because of the computational cost of generating a training set of \EgHSE{} values is exorbitantly high.
\end{itemize}

In the third step of the workflow outlined above, the key criteria in selecting the materials is that they are stable based on \EfPBEML{}. Additionally, we only considered compounds composed of chemically reasonable elements, excluding $f$-block elements such as lanthanide-containing compounds because of the computational expense and limitations of DFT in describing elements with $f$-electrons,\cite{zhang2020chemical} which is discussed in more detail below. 

\subsection{Screening of unseen AA'BB'X$_{6}$ compounds using the ML-predicted formation energy} \label{sssec:secdHf}

The specific set of samples used in the training set directly impacts the accuracy of the ML model. Often, this selection process becomes a computational bottleneck because labeling the dataset can be expensive and time-consuming. To address this we used an active learning approach, rather than randomly selecting samples for the training set. This method allows the model to selectively query the most informative samples to the training set based on uncertainties. By doing so, we optimize the learning process to improve model accuracy more efficiently and effectively while minimizing the need for costly labeling.

\begin{figure}[htbp!]
  \centering
  \includegraphics[width=0.5\textwidth]{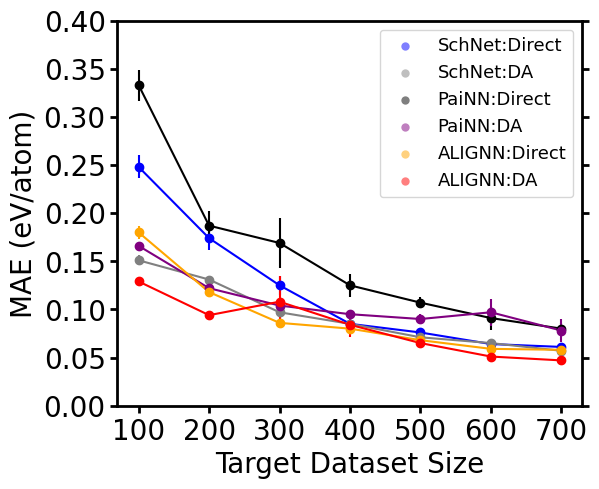}
  \caption{Change in the test-set MAE of \EfPBEML{} (compared with \EfPBE{}) calculated on a fixed test set of 270 samples, with the training set size increasing in increments of 100 samples. In all cases, \sourceABXD{} ABX$_{3}$ samples were used as the source dataset.}
  \label{fig:scaling_plot}
\end{figure}

We initiated the training of \EfPBEML{} using 970 (out of \entireD{}) randomly selected AA'BB'X$_{6}$ compounds. For a comparison of the performance of DA and direct training procedures at small-dataset sizes, the scaling curves were generated by incrementally increasing the target dataset size from 100 to \iterI{} samples, while keeping a fixed 270-sample test-set (\cref{fig:scaling_plot}). For the direct training strategy, all three architectures were initialized with a random set of weights. At small dataset sizes indicates that the DA models consistently outperform the directly trained models for all three architectures, with a lower standard deviation across three independent training iterations. However, as the number of target samples increased, the MAEs of both the DA and direct training models converged. For \Sch{} and \AL{}, this convergence occurred with 400 training samples, while for \PN{}, convergence was reached between 400 to 600 samples. Although a substantial improvement is observed when the number of samples in the target dataset is increased, increasing the source dataset size had little effect on the retraining of the target model (see Fig. S.1). 

These results indicate that the DA scheme effectively transfers knowledge of the chemical environments from the source dataset to the training of the target dataset. In this work, a dataset of \sourceABXD{} ABX$_{3}$ compounds was used as the source dataset. We tested how the choice of the source dataset affects the model performance, and found that using the source dataset of \sourceABXD{} ABX$_{3}$ halides provides a considerably more accurate model than using the DA scheme with a source dataset of AA'BB'O$_6$ (i.e. oxides) compounds ( see Fig. S.2). Thus, it appears advantageous that the source dataset have a similar chemical environment to that of the target dataset. 

\begin{figure}[htbp!]
\centering
    \captionsetup[subfigure]{labelformat=empty}
    \begin{subfigure}[b]{0.5\textwidth}
        \includegraphics[scale=0.5]{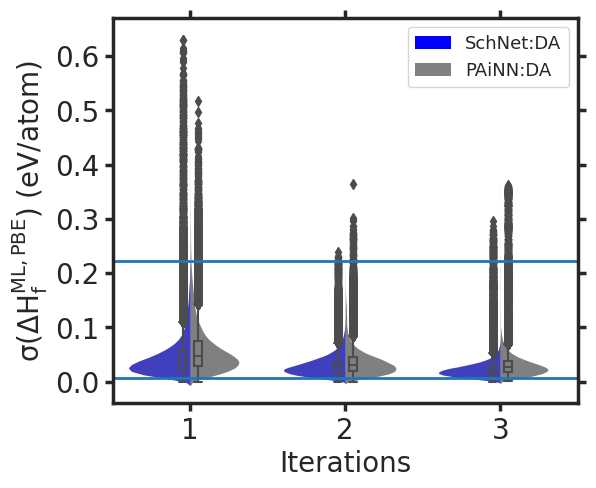}

        \caption{(a)}
    \end{subfigure}
    
    \begin{subfigure}[b]{0.5\textwidth}
        \includegraphics[scale=0.5]{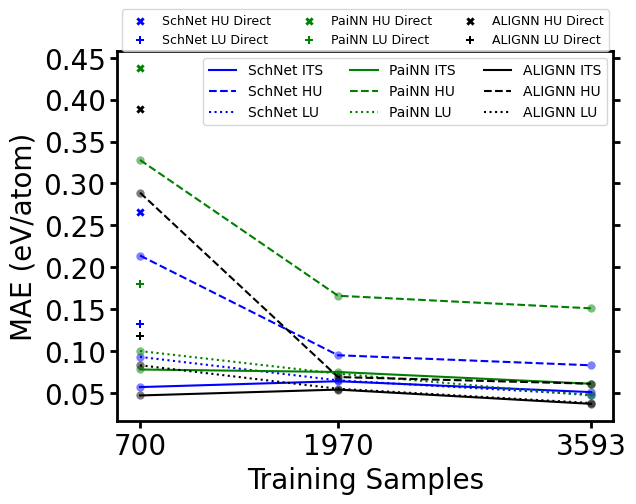}

         \caption{(b)}
    \end{subfigure}
    
\caption{ (a) Distribution of the standard deviation across three independently trained \EfPBEML{} models over unseen samples, denoted $\sigma$(\EfPBEML), using DA \PN{} and \Sch{}; (b) Evolution of MAE for predicted \EfPBE over three active learning iterations. The number of training samples in Iteration 1, 2 and 3 are 700, 1970, and 3593, respectively. The number of initial test-set (ITS), high uncertainty (HU), and low uncertainty (LU) samples in Iteration 1 are 270, 1679, and 1873, respectively. In Iteration 2, these numbers are 1873 (ITS), 1623 (HU), and 1877 (LU). In Iteration 3, the numbers are 1877 (ITS), 790 (HU), and 948 (LU), respectively.}
\label{fig:hf_mae_violin_uncertainty}
\end{figure}

The initial models were trained using \iterI{} training samples, referred to as Iteration 1 in \cref{fig:hf_mae_violin_uncertainty}A, and then applied to an unseen set of the remaining \screenI{} AA'BB'X$_{6}$ compounds. Instead of randomly selecting samples from the extensive dataset of unseen compounds, these were ranked based on their uncertainty. This ranking helped identify which compounds should be added to the training set in subsequent iterations to improve ML model performance. Uncertainties for each sample were calculated as the standard deviation, denoted as $\sigma$, across predictions from three independently trained ML models using the DA approach.

We evaluated the model errors using the excluded samples and divided them into high-uncertainty (HU) and low-uncertainty (LU) samples based on the magnitude of their uncertainties (i.e., $\sigma$). \cref{fig:hf_mae_violin_uncertainty}A shows the range of $\sigma$ values at each iteration of the active learning loop. The selection criteria for the HU and LU samples were based on the 99\textsuperscript{th} and 1\textsuperscript{st} percentiles of $\sigma$ values, respectively, at each iteration. Note that for the HU samples, all Nd-containing samples were excluded because of their consistently higher uncertainties (see Fig. S.6) and due to fact that Nd-containing compounds are not of interest for PV applications. Furthermore, in the next section, we discuss the justification for excluding $f$-block elements from the training set for \EgHSE{}.

For the initial model trained to \iterI{} samples, much larger errors for the HU and LU samples compared with the initial test-set errors evaluated for the 270 samples (\cref{fig:hf_mae_violin_uncertainty}b) were observed for all three models, indicating poor model generalizability. Nonetheless, for these two sets of unseen samples, the DA models all had substantially lower errors compared to the direct models by factors of 1.35 (1.80), 1.24 (1.42), and 1.35 (1.42) for the HU (LU) samples for \PN, \Sch, and \AL{} models, respectively (see \cref{fig:hf_mae_violin_uncertainty} b and Table S.3). 

Despite these quantitative differences, it is interesting to note that the direct and DA models appear remarkably indistinguishable based on test set errors. The MAEs are 61 / 57 meV/atom, 80 / 78 meV/atom, and 58 / 47 meV/atom for direct / DA for \Sch, \PN, and \AL, respectively, for the 270-sample test set used in Iteration 1. In addition to comparable accuracies on the same test set, both models show a high correlation in the ML-predicted \Ef{} for the \iterI{} unlabeled compounds in Iteration 1 (R$^{2}$ = 0.949 and 0.974, respectively, see Figs. S.3a and S.3b). 

Three active learning loops (i.e., samples are ranked based on uncertainties and then the samples with the highest uncertainties are added back to the training set and the models are retrained) were performed until a similar error was achieved between the unseen HU and LU samples and the initial test-set. These final models (referred to as Iteration 3) were re-trained using \iterIII{} samples (1623 HU samples selected from the Iteration 2 model were added to the \iterII{} training samples used to train Iteration 2)(see Table S.3). In general, these results indicate that $\approx 1~\%$ of the samples from the overall space, is sufficient to accurately predict the \EfPBE{} of the entire space. This results in a substantial reduction in computational resources and labeling efforts, demonstrating the effectiveness of active learning in optimizing the training set composition for maximum accuracy with minimal data.

For \AL{}, MAEs of 61 and 38 meV/atom were observed for 790 HU and 948 LU samples, respectively, which is comparable to the initial test-set (MAE = 37 meV/atom). Compared with \AL{}, larger differences were calculated in the MAEs of the HU and LU samples (MAE = 83/48 meV/atom for HU/LU, respectively) for \Sch{}, and even more so for \PN{} (MAE = 151/47 meV/atom, respectively) compared with their test-set MAEs of 51 and 61 meV/atom, respectively (see Table S.3). Although the errors for \Sch{} and \AL{} became acceptable, the significant differences in the performance of the \PN{} model between the HU and LU samples suggest a potentially optimistically biased model. Due to the accuracy of the model with the left-out HU and LU samples, the \AL{} model was selected for screening stable new compounds.

\begin{figure*}
\centering
    \captionsetup[subfigure]{labelformat=empty}
    \begin{subfigure}[b]{0.33\textwidth}
                \includegraphics[scale=0.6]{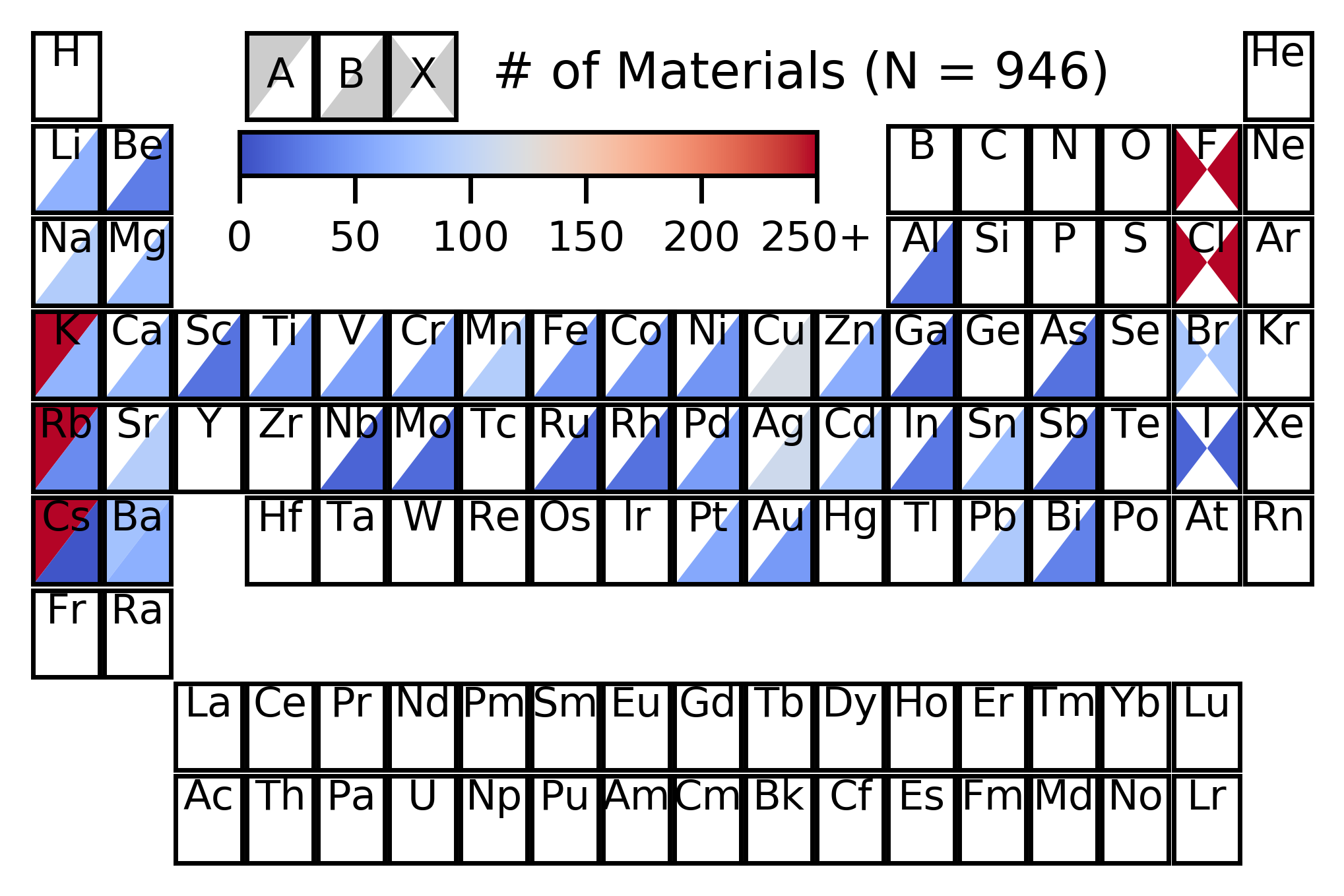}
                \caption{(a)}
    \end{subfigure}
        \begin{subfigure}[b]{0.3\textwidth}
                \includegraphics[scale=0.35]{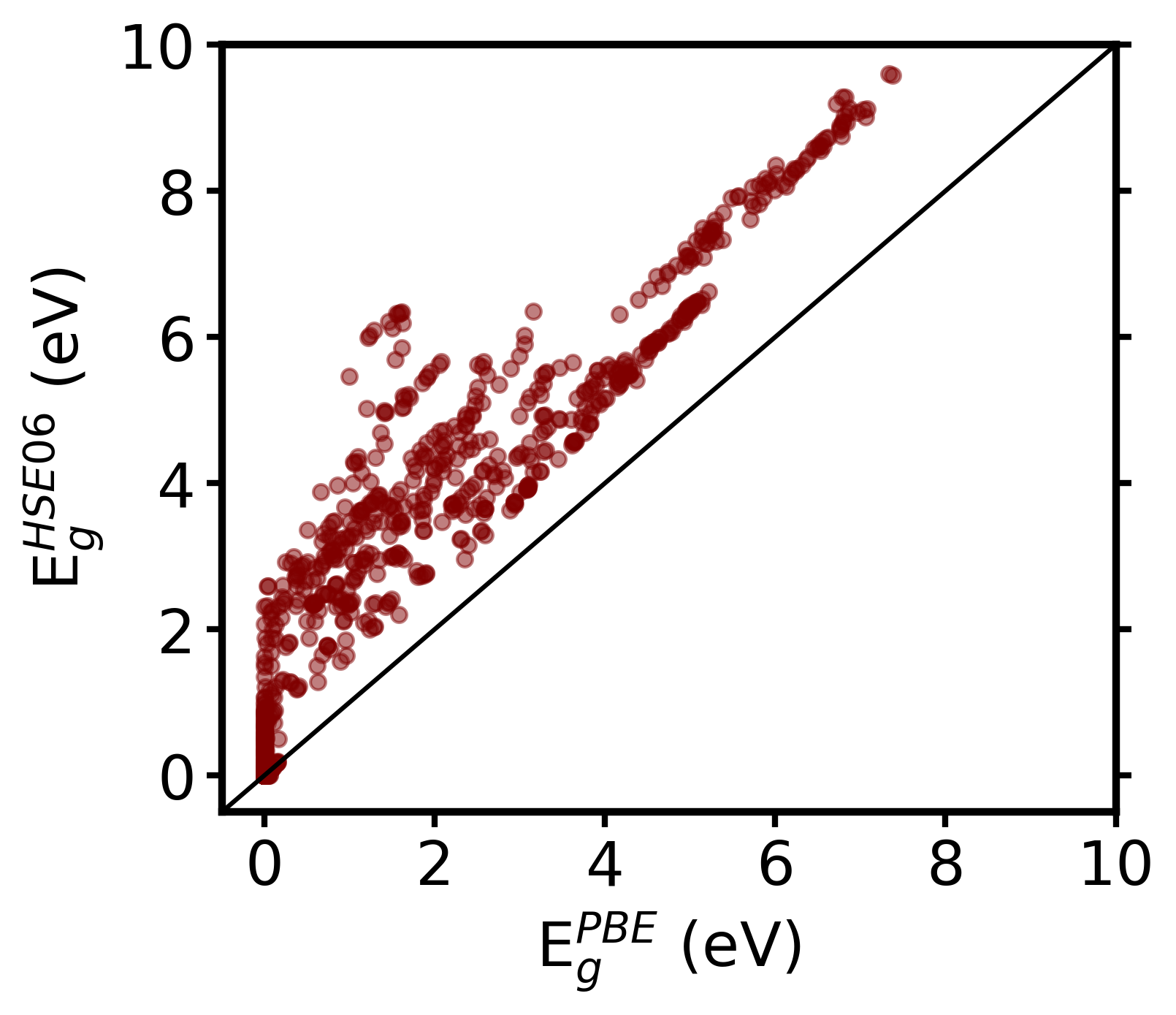}
                 \caption{(b)}
    \end{subfigure}
        \begin{subfigure}[b]{0.3\textwidth}
                \includegraphics[scale=0.35]{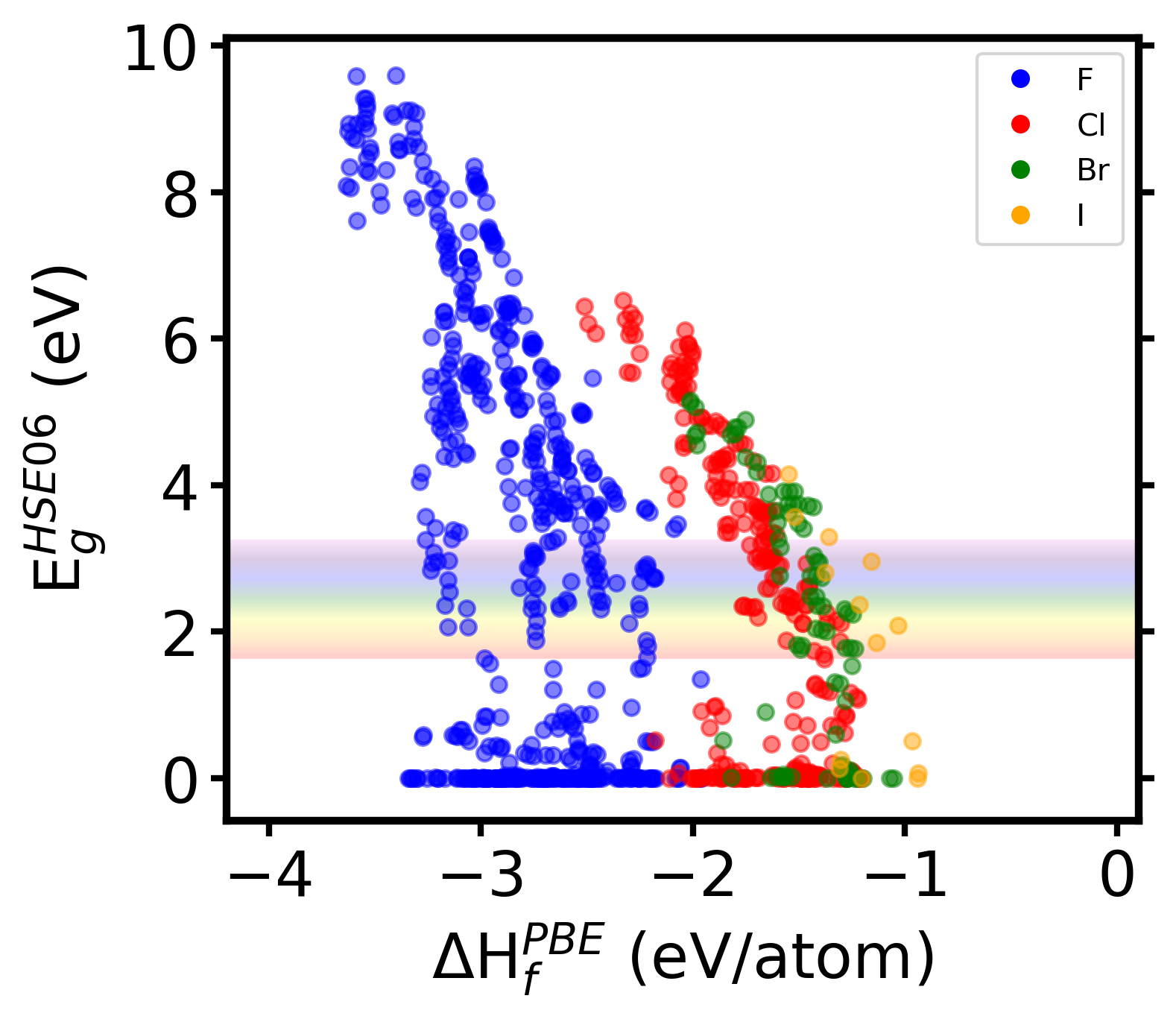}
    \caption{(c)}
    \end{subfigure}

    \caption{(a) Periodic table plot showing the A-site (upper left corner), B-site (lower right corner), and X-site (left and right triangles) frequencies (heatmap) of each element present in our subset of \entireDEg{} samples for which \EgHSE{} was computed. (b) Parity plot showing that \EgPBE{} consistently underestimates \EgHSE. (c) Distribution of calculated \EgHSE{} and \EfPBE.}
       \label{fig:subset}
\end{figure*}

\subsection{Developing a machine learning model for predicting PBE and HSE band gap energies} \label{secEg}

\begin{table}[htbp!]
\centering
\caption{Comparison of MAEs (in eV) for \EgPBEML{} and \EgHSEML{} for a left-out test set of \NtestEg{} samples. All models were trained on \NtrainEg{} samples. In the DA procedure, each model was initially trained to \EgHSE{} and \EgPBE{} on a source set of \sourceABXDEg{} and \sourceABXD{} \ce{ABX3} compounds respectively. In the transfer learning (TL) scheme, \EgHSEML{} was initialized using the \EgPBEML{} model trained to \NtrainEg{} AA'BB'X$_6$ samples.}
\label{tab:mae_bandgap}
\resizebox{\columnwidth}{!}{%
\begin{tabular}{|c|p{1.2cm}|p{1cm}|p{1cm}|p{1cm}|p{1cm}|}
\hline
\textbf{Model} & \textbf{Proce-dure} & \multicolumn{2}{c|}{\EgHSE{} (eV)} & \multicolumn{2}{c|}{\EgPBE{} (eV)} \\ \hline
               &                            & \textbf{BVM}     & \textbf{DFT}     & \textbf{BVM}     & \textbf{DFT} \\ \hline
\Sch{}         & \centering\arraybackslash TL  & 0.190  &  0.214 & - & -       \\ \cline{2-6} 
                & \centering\arraybackslash DA  & 0.181  &  0.177 & 0.140 & 0.106       \\ \cline{2-6}
               & \centering\arraybackslash Direct & 0.245 & 0.226 & 0.158 & 0.128       \\ \hline
\PN{}          & \centering\arraybackslash TL  & 0.267 & 0.219 & - & -       \\ \cline{2-6} 
                & \centering\arraybackslash DA  & 0.236 & 0.203 & 0.122 & 0.125       \\ \cline{2-6}
               & \centering\arraybackslash Direct & 0.276 & 0.327 & 0.172 & 0.144       \\ \hline
\AL{}          & \centering\arraybackslash TL  & 0.214 & 0.155 & - & -       \\ \cline{2-6} 
                & \centering\arraybackslash DA  & 0.188 & 0.166 & 0.109 & 0.099       \\ \cline{2-6}
               & \centering\arraybackslash Direct & 0.227 & 0.170  & 0.116 & 0.108       \\ \hline
\end{tabular}
}

\end{table}

As described in the section \ref{sssec:dcp}, we chose a subset of compounds that demonstrated stability according to the ML-predicted \EfPBE{} from Iteration 1 for which \EgHSE{} was calculated and \EgHSEML was trained. The subset of \entireDEg{} most frequently have K, Rb, Cs, and Ba on the A-site because of the higher likelihood for forming stable compounds when these elements occupy the A-site. Larger elements from Group 1 (Cs, Rb, K, Na) and Group 2 (Ba, Sr) will most likely occupy the A-site, which can be observed from the difference in \EfPBEML{} for compounds containing these cations on the A-site compared with the B site (Fig. S.5). In separate work, these elements have also been reported to enhance the stability of perovskites, partially because of their relatively large ionic radii.\cite{zhang2020chemical} Despite the training/test set being limited to only \entireDEg{} compounds and predominantly consisting of compounds with only a few different elements on the A-site, the elements on the B and B' sites in this set span 35 different elements, despite excluding the lanthanides. This chemical diversity enables an investigation of how variations in the nature of the B-X bond affect the electronic properties. \cref{fig:subset}C compares the range of \Ef{} and \EgHSE{}, with these compounds spanning over a 9 eV range in \EgHSE{}. For this subset, the mean absolute error between \EgPBE{} and \EgHSE{} is 1.21 eV. 

The most accurate models relative to \EgHSE{}, are consistently based on the DA-scheme, which for \Sch{} and \AL{} results in excellent accuracies of MAE = 181 meV and 188 meV using the BVM structure, with slightly higher errors are observed for \PN{} (236 meV). Slightly higher accuracies can be achieved when the PBE structure is used instead, but the small improvement in accuracy does not justify the considerable additional expense of performing a PBE optimization. Using the same training/test set split, the DA approach also leads to an improvement over the direct approach for \EgPBE{} for \Sch{} (140 meV vs. 158 meV for DA and direct, respectively) \PN{} (122 meV vs. 172 meV) and \AL{} (109 meV vs. 116 meV) for the excluded test set (see Table \ref{tab:mae_bandgap}).

Because of the computational cost of generating a source dataset of \EgHSE{}, in addition to DA, transfer learning (TL), \cite{weiss2016survey, yamada2019predicting, jha2019enhancing} was also used by leveraging an initial model was trained on \EgPBE{} for \NtrainEg{} AA'BB'X$_6$ samples. In TL, a separate model was trained but initialized with the parameters from the \EgPBE-trained model and fine-tuned for \EgHSE{} (using the same \NtrainEg{} AA'BB'X$_6$ samples). The performance of these models is then assessed using the same \NtestEg{} AA'BB'X$_6$ compounds. The TL approach resulted in MAEs for \EgHSE{} of 190, 267 and 214 meV, for \Sch{}, \PN{}, and \AL{}, respectively on the test set of \NtestEg{} compounds using the BVM structure as input. In comparison, the direct training approach led to MAEs of 245, 276, and 227 meV, respectively, indicating that the TL approach led to an improvement for all three models (Table \ref{tab:mae_bandgap}). Training the ML model to the PBE structure instead leads to similar results, indicating that the BVM structure, which is an inexpensive proxy structure, is a significant advantage in screening these materials.  

We note that an alternative scheme for predicting \EgHSE{} is $\Delta$-learning, which was previously used by some of us to learn the HSE06 quality band structures.\cite{adhikari2023accurate} However, $\Delta$-learning requires inputting \EgPBE{} to predict \EgHSE{} -- i.e., requires a successful PBE geometry optimization prior to making a prediction -- which is more computationally demanding than the DA/TL schemes that make it possible to predict \EgHSE{} directly from the BVM structure. 

Separately, the subset of \entireDEg{} compounds were selected from Iteration 1 and not used in subsequent active learning iterations, and therefore, represent an additional totally left-out set for evaluating the performance of ML models from later iterations in predicting \EfPBE{} to assess the generalizability of the ML models. This subset shows a different distribution of \EfPBE{} compared to  training samples in Iteration 1 (see Fig. S.4) indicating that they represent an out-of-domain distribution. As a result, the DA-based models trained in Iteration 1 had high errors for \EfPBE{} for these samples specifically (MAE = 107, 159, and 142 meV/atom for \Sch{}, \PN{}, and \AL{}, respectively) ( see Table S.2). However, these errors were factors of 2.0, 1.9, and 1.7 lower than the directly-trained ML models from Iteration 1 (MAE = 215, 208, and 244 meV/atom, respectively). The higher accuracies for the DA-approach indicate an enhancement in the ML model's generalizability when trained through the DA approach. By Iteration 3, the accuracies of the DA-models improved significantly to MAE = 45, 56, and 34 meV/atom for \Sch{}, \PN{}, and \AL{}, respectively (as shown in Table S.2). This trend aligns well with the errors observed for the excluded HU and LU samples, which also showed that sufficient accuracy in \EfPBEML{} could be obtained by Iteration 3.

\subsection{Screening materials based on \EfPBEML{} and \EgHSEML}

\begin{figure}[htbp!]
\includegraphics[scale=0.60]{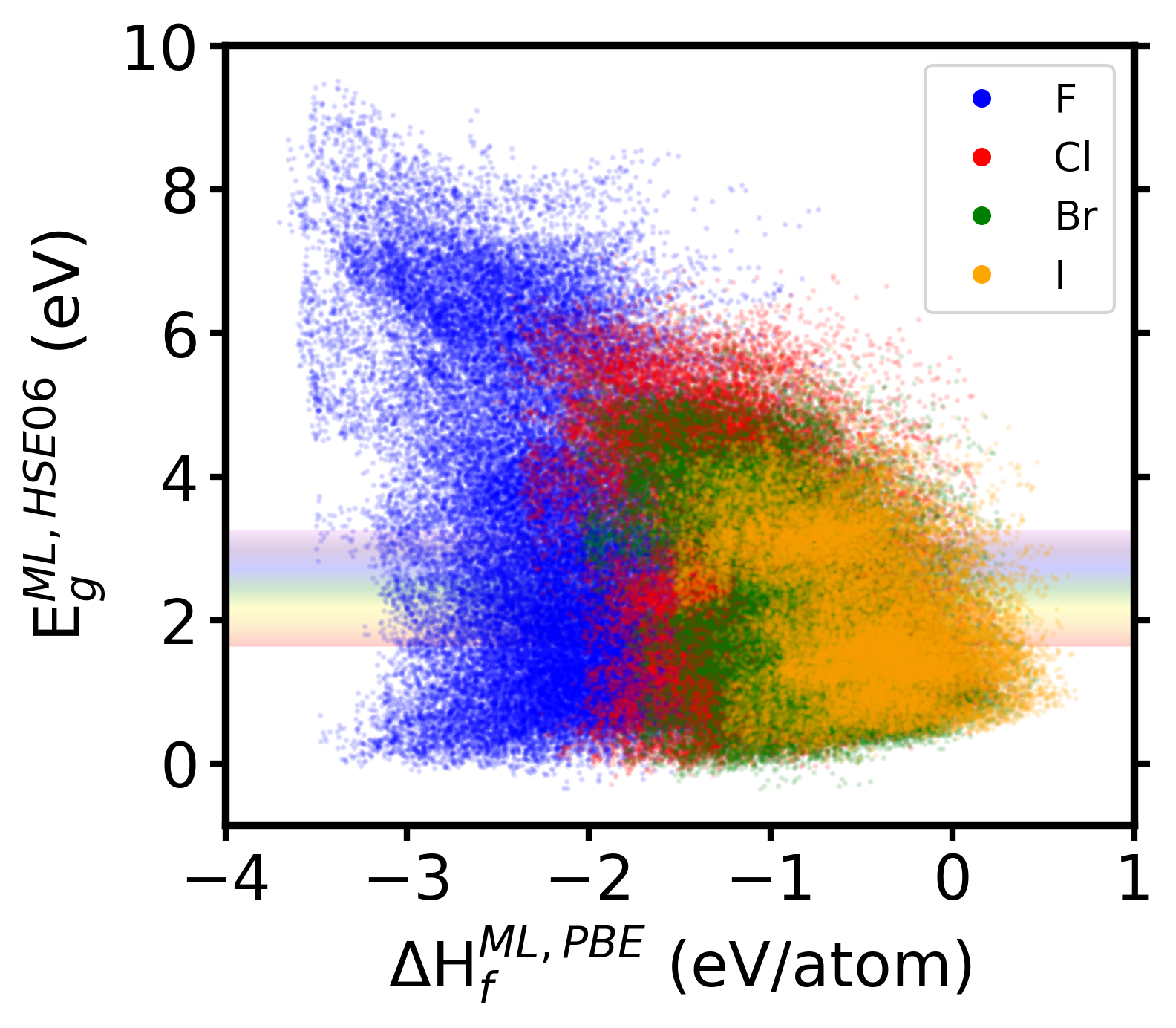}
\caption{ \label{fig:v3 screening} The DA \AL{} model \EfPBEML{} and \EgHSEML{} for \entireD{} perovskite compositions. Compounds with $X$ = I tend to possess higher \Ef{} and smaller \Eg{} whereas compounds with $X$ = F tend to have the lowest \EfPBEML{} and largest spread of \Eg{}. Compounds with $X$ = Br, Cl tend to have intermediate \Ef{} and \Eg{} with Cl possessing the larger \Eg{} on average.}
\end{figure}

The DA \AL{} models trained to predict \EfPBEML{} (description in Section \ref{sssec:secdHf}) and \EgHSEML{} (description in Section \ref{secEg}) for an unseen set of \screenIII{} compositions as well as adding the compounds from the training set, leading to evaluation of properties for \entireD{} samples in \cref{fig:v3 screening}. Note that these predictions \textit{include} compounds containing $f$-block elements, even though DFT struggles with these elements and they were intentionally excluded from the training set for \EgHSEML{}. Therefore, in these cases, the predictions are expected to be unreliable, yet they were still included (but are labeled as such in Table S.6).  

The \EfPBEML{} values are negative indicating stability against the competing elemental phases, except for a few cases that generally contain $X$ = I. Compounds that contain $X$ = F tend to exhibit the lowest \EfPBEML{} values followed by $X$ = Cl, Br, and I. \EfPBEML{} is used to compute \EdPBEML through the convex hull construction using all competing phases available in OQMD (see Methods section for more details). Out of the \entireD{} compositions, only $\approx2~\%$ (3410 samples) are predicted to be stable/meta-stable with \EdPBEML{} < 60 meV/atom. 

Regarding band gaps, we selected a range of 1.0 eV $<$ \EgHSEML{} $<$ 1.6 eV for the screening process. This range accounts for the Shockley–Queisser limit of 1.34 eV, but expanded by $\pm$ 0.3 eV because of the error in the \EgHSEML{}. Of the \entireD{} compounds evaluated (i.e., all training and test set samples were considered in this screening), only \subsetD{} candidate compounds are within the range of 1.0 eV $<$ \EgHSEML{} $<$ 1.6 eV and are relatively stable with a \EdPBEML{} $<$ 60 meV/atom. Note: this set excludes 69 $f$-block compounds that fall within this range but are considered unreliable. Moreover, an additional 21 compounds with Tl on the A-site were excluded because of the toxicity of Tl. 

These \subsetD{} candidates are a chemically diverse set, containing a total of 6 elements on the A-sites and 17 elements on the B-sites. The majority of candidates predominantly have 3 elements on the A-sites (Cs, Rb, K) and 3 elements on the B-sites (Cu, In, Rh). 67~\% of the \subsetD{} candidate materials contain either Cu, In, or Rh on the B-site. Regarding the X-site, the range and magnitude of \EgHSEML{} increases as the radius of X decreases from I to Br, Cl and F. This agrees with previous computational results, where stability and \Eg{} tends to increase as X becomes smaller (from I to F).\cite{zhang2020chemical, im2019identifying} Because of the general trend towards increasing band gap energies in substituting anions from I to F, the set of \subsetD{} candidates contains Cl (50~\%) most frequently, followed by Br (27~\%), and then F (23~\%). 

To verify the accuracy of the model, we randomly selected 8 compounds from the set of \subsetD{} and subsequently calculated \EgHSE{}. These errors are reported for each of the 8 compounds in Table S.5 and show an MAE = 0.14 eV. The projected density of states for 6 of these 8 compounds is displayed in Fig S.7. Moreover, several of the \subsetD{} candidates have been previously investigated for their potential in PV applications. These prior studies provide a useful basis for comparison when assessing the accuracy of \EgHSEML{} against independent results. Included in this set of \subsetD{} candidates are five compounds present in the training data set (\ce{RbCsAgInBr6}, \ce{Cs2GaAgBr6}, \ce{Cs2AgInBr6}, \ce{KRbCuRhF6}, \ce{KCsCuInCl6}), two of which have been already investigated, \ce{Cs2GaAgBr6}\cite{zhao2017cu} and \ce{Cs2AgInBr6},\cite{zhao2017cu} and as expected compare well with \EgHSEML (see Table \ref{tab:pv_candidates}). Six compounds in the test set have been previously investigated: \ce{Cs2CuSbCl6},\cite{zhou2020lead} \ce{Rb2CuInCl6},\cite{zhao2017cu} \ce{Rb2AgInBr6},\cite{zhao2017cu} \ce{K2CuGaCl6},\cite{zhao2017cu} \ce{K2CuInCl6},\cite{zhao2017cu} and \ce{Rb2GaAgBr6},\cite{zhao2017cu} and their reported \EgHSE{} values are on average signed difference of -0.14 eV (indicating consistently underestimating other calculated results) with the largest difference of -0.31 eV of \EgHSEML{} for \ce{Rb2CuInCl6}. Finally, four of the \subsetD{} candidate materials identified here were previously disregarded. \ce{Cs2CuInCl6}, \ce{Cs2GaAgBr6}, \ce{K2CuInCl6}, and \ce{K2CuGaCl6} failed the criteria for stability with respect to binary competing phases.\cite{zhao2017cu} Overall, the strong agreement between these ML predictions and the independently reported values for these compounds — 4 of which have been reported in the literature as being stable  — further emphasizes the potential of the \subsetD{} candidate compounds identified in this study.

\begin{table}[ht]
\centering
\caption{Comparison of the \EgHSEML{} for samples selected from \entireDEg{} against those with previously reported \EgHSE{} values. All values are in eV}
\label{tab:pv_candidates}
\resizebox{\columnwidth}{!}
{%
\begin{tabular}{|c|c|c|c|c|}
\hline
\textbf{Compound} & \textbf{Dataset} & \textbf{Exp. \Eg (eV)} & \textbf{\EgHSE (eV)} & \textbf{\EgHSEML (eV)} \\ \hline
\ce{Cs2GaAgBr6} & Training & - & 1.32  & 1.49 \\ \hline
\ce{Cs2AgInBr6} & Training & 1.56  & 1.25-1.50 & 1.50  \\ \hline
\ce{Cs2CuSbCl6} & Test & 1.66 & 1.70  & 1.56 \\ \hline
\ce{Rb2CuInCl6} & Test & - & 1.36  & 1.05  \\ \hline
\ce{Rb2AgInBr6} & Test & - & 1.46 & 1.39 \\ \hline
\ce{K2CuGaCl6} & Test & - & 1.30  & 1.29 \\ \hline
\ce{Cs2CuInCl6} & Test & - & 1.40 & 1.24 \\ \hline
\ce{K2CuInCl6} & Test & - & 1.35 & 1.08 \\ \hline
\end{tabular}
}

\end{table}

\section{Methods}

To create a system-specific structure for each of the \entireD{} compounds, the optimal lattice parameters for the 10-atom unit cell were obtained by minimizing the Global Instability Index (GII) for each composition. GII quantifies the deviation from ideal bond valences and serves as an indicator of structural stability. To calculate bond valences, oxidation state-specific cation-anion bond valence parameters from the 2020 update by Brown et al.,\cite{bvp} and only cation-anion bonds less than 6 \AA{} were considered. The minimization of GII for structure generation has previously been shown to accurately reproduce experimental \cite{lufaso2001prediction} and theoretical \cite{morelock2022bond} perovskite structures on a large scale.\cite{bare2023dataset} The structure prediction code is publicly available at: https://github.com/zaba1157/bvm\_cubic\_perovskite. The same GII minimization scheme was used to generate the primitive structures for all \entiresource{} single perovskites with the formula \ce{ABX3}, all within the cubic Pm-3m space group (\#221). 

Band gap energies were computed using a fully converged single self-consistent field calculation on the PBE-optimized geometries with both the PBE and HSE06 functionals (simply referred to as HSE). The tetrahedron method with Bl{\"o}chl corrections and a $\Gamma$-centered k-mesh were used for all band gap calculations. All DFT calculations were performed within the PAW formalism as implemented in the Vienna ab initio simulation package (VASP) version 5.4.4.\cite{kresse1994norm, kresse1996efficiency, kresse1996efficient} These PBE geometry optimizations were performed with a plane-wave cutoff energy of 600 eV and an energy convergence criterion of $10^{-6}$ eV. Gaussian thermal smearing with a parameter $k_{B}T$ of 0.02 eV was used in all calculations. All magnetic systems were initialized in a ferromagnetic spin state. $\Gamma-$centered k-meshes were used with the smallest allowed spacing between k-points of 0.1 \AA{}$^{-1}$ for the Brillouin zone sampling.

\Sch{}~\cite{schutt2018schnet} is a continuous-filter convolution neural network that encodes information from the atomic local environments by first determining the neighboring atoms within an 8 \AA{} cut-off distance. Then, to encode many-body interactions from these local environments, message passing is performed from three key components: atom-centric layers, interaction modules, and filter generation networks. 

The atom-centric layers perform transformations on atomic embeddings, derived from atomic numbers, using multiple dense layers with each layer having a width of 128. \Sch's convolution module updates edge features through localized pooling, integrating the information from neighboring nodes. The interaction module further combines node and edge features through three dense layers and an addition module, with each layer again having a width of 128 and employing a shifted softplus function for non-linear activation. Downstream predictive tasks leverage a multi-layer perceptron with several dense and pooling layers. The network's depth was consistently set to eight for all experiments.

\PN{} is an equivariant neural network that captures the direction of interatomic distances using directional vectors, in contrast to the scalar representation of distances used by models such as \Sch{}. It transforms the Euclidean distances between atom pairs into an equivariant representation using a Bessel basis with 20 radial functions, optimized for a 3 \AA{} cutoff. We tested larger cutoffs of up to 6 \AA{} without significant performance gains. Each radial function passes through a series of layers—input, hidden, and output—to calculate atomic interaction weights. For convolution, \PN{} applies a sequence of three dense layers, paired with two pooling and two gathering layers to integrate invariant and equivariant features—a process executed via the Keras library \cite{REISER2021100095}. This is followed by an update block that refines both node and equivariant features, utilizing four dense layers along with various geometric and arithmetic modules.

\AL{} combines a bond graph from interatomic distances with a line graph.\cite{choudhary2021atomistic} The nodes of the bond graphs correspond to atoms and its edges correspond to bonds while the nodes of a line graph correspond to interatomic bonds and its edges correspond to bond angles. The inclusion of bond distance and bond angle has incorporated finer details of atomic structure that leads to higher model performance predicting solid state and molecular properties.\cite{choudhary2021atomistic} \AL{} performs edge gated graph convolution where node features and edge features are updated concurrently. The node feature set is the combination of 9 inputs, which is inspired by the crystal graph convolutional neural network (CGCNN) model.\cite{xie2018crystal} The initial edge features are interatomic bond distances further expanded via radial basis functions. Each \AL{} layer consists of edge-gated graph convolution over the line graph followed by another edge gated graph convolution over the bond graph. The first convolution produces the bond messages that are used by the next edge-gated graph convolution module to update the node feature and bond features of the original graph. The depth and width of the model have been set as 4 and 256, respectively, with an 8 \AA  \ cut-off distance. A sigmoid linear activation function has been used for its twice differentiability and better empirical performance. After graph convolution, global pooling was performed over the nodes, and the final prediction was made with fully connected layers. 

\begin{figure*}[htbp!]
      \centering
      \includegraphics[width=0.80\textwidth]{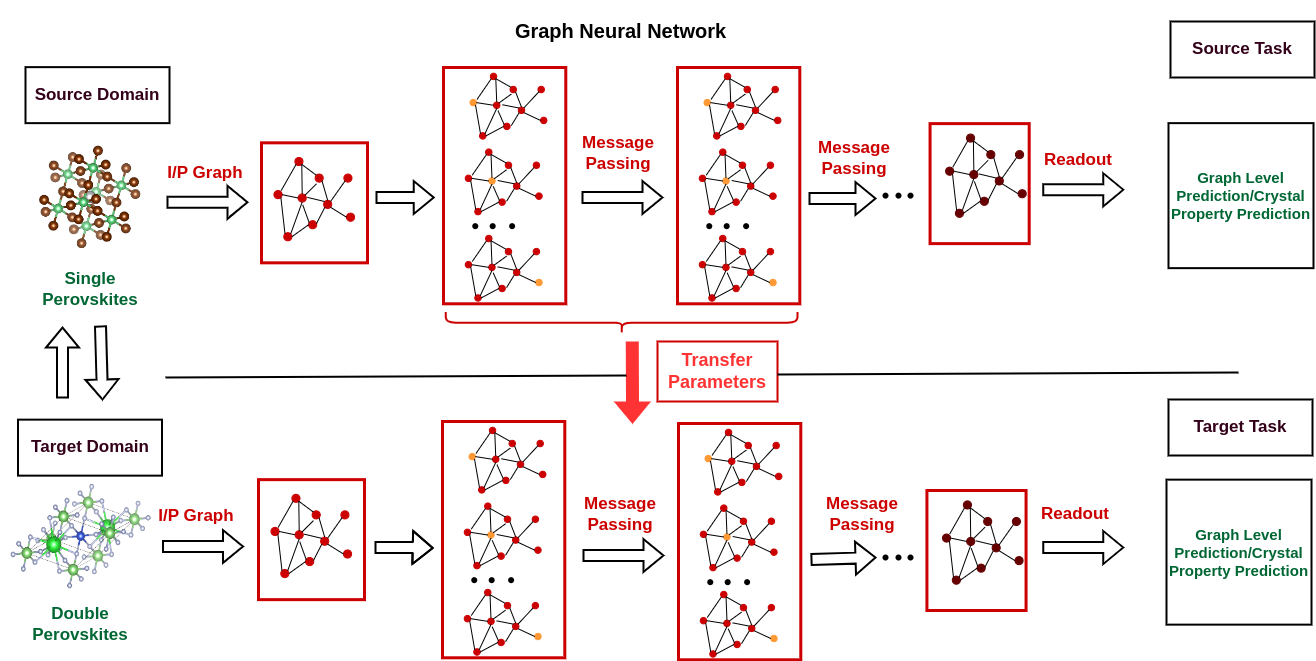}
       \caption{The DA approach involves two distinct pipelines, each with a different dataset: source and target. For the source dataset, \sourceABXD{} ABX${3}$ compounds were used to train an initial model. In the DA scheme, the optimal parameters/weights from the source dataset training are transferred to begin training a new ML model on the target dataset. This model is then further refined through standard validation and testing procedures. The final target model is used to predict the properties of a new set of \entireD{} AA'BB'X${6}$ compounds.}
      \label{fig:param_da_pipeline}
\end{figure*}

\section{Domain Adaptation}
DA is the task of improving the performance of a model on a target domain containing insufficient data by using the knowledge learned by the model from another related domain. In the pursuit of building a good estimator for the target domain, several DA strategies have been developed over the past few years,\cite{farahani2021brief, de2021adapt} with different objectives, such as: 
  \begin{itemize}
      \item \textbf{Feature-Based DA} works with an encoded feature space to correct the difference between source and target distributions.
      \item \textbf{Instance-Based DA} re-weights training data to correct the difference between source and target distributions.
      \item \textbf{Parameter-Based DA} passes parameter/weights from the pre-trained model to the target model to build a suited model for the target task.
  \end{itemize}

Because generating a DFT-calculated training dataset for compounds with a large number of atoms is computationally demanding, our goal is to recycle the learnable parameters and weights that is learned in the source training and apply it to new systems. To do this, we used the parameter-based DA approach. Feature-based DA requires a new encoding mechanism or a labeling function for the target domain while instance selection and instance weighting is followed by the instance based DA to learn the importance of labeled data.\cite{instda} However, due to having limited source and target data, we decided to use the data unmodified. Therefore, we worked on the model level specifically with the weights and parameters rather than the dataset to deploy DA. The parameter-based DA approach, which is illustrated in \cref{fig:param_da_pipeline}, first involves training an ML model on an initial dataset (called the source dataset), then the parameters of this initial model are used to initialize a subsequent ML model for training to the target dataset. Our specific application of GNN architectures of materials involves transferring weights and parameters of different layers learned on the source dataset to a target dataset.

The source ML model was trained on the source dataset for a total of 250 epochs, with 50 samples per batch. At each step within an epoch, the error of the model predictions were calculated against the actual labels for these 50 samples. The optimal parameters and weights from the initial training were used to initialize the retraining on the target dataset. A learning rate of 0.0005 was used across all training of the source and target models and the MAE was used as the evaluation metric. We have also tested the learning rates of 0.005 and 0.00005 for the \Sch{} and \PN{} DA training. Similar performance has been observed for \Sch{} while the maximum difference for \PN{} was 14 meV/atom. An end-to-end training of the source dataset and retraining of the target dataset was performed three times, yielding three distinct models.

\section{Conclusions}

In this work, we demonstrate that domain adaptation (DA) improves the performance of ML models consistently across three different architectures applied to three different target properties of \EfPBE{}, \EgPBE, and \EgHSE. DA allows existing models to be efficiently fine-tuned for new domains or tasks, significantly reducing the resources and data typically required for training robust ML models from scratch. Indeed, combining the DA scheme with active learning, high accuracies were achieved in the prediction of \EfPBE{} (MAE = \ALerrEf{}) and\EgHSE{} (MAE = \ALerrEg{}) using training sets with a relatively small number of compounds. We also observed that the DA scheme improved the generalization of the ML models. To our knowledge, these results are the first demonstration of the DA scheme for GNNs to predict material properties.  

Moreover, these accurate models allowed for screening a large space. First, \EfPBEML{} was used to compute the decomposition enthalpy (\EdPBEML{}) through convex hull construction for all \entireD{} compounds, which utilized all competing phases available in OQMD. Of the \entireD{} compositions analyzed based on \EdPBEML{}, only $\approx2~\%$ (3410 compositions) are predicted to be (meta)stable with \EdPBEML{} < 60 meV/atom. This set was further reduced when considering the band gap energies of the material. Of the \entireD{} compounds, only \subsetD{} candidate compounds fall within the desired range of 1.0 eV $<$ \EgHSEML{} $<$ 1.6 eV and display \EdPBEML{} $<$ 60 meV/atom. Four of these were previously disregarded due to instability and 4 have been previously identified as promising for PV applications, which aligns well with our ML model predictions. The remaining dataset of 40 compounds remains unexplored to the best of our knowledge and therefore offer opportunities for potential experimental examination. 


\section*{Acknowledgement}
We gratefully acknowledge funding from the Department of Energy Office of Energy Efficiency and Renewable Energy within the Solar Energy Technologies Office through award number DE-EE0009515. Calculations were carried out on the computational resources located at the National Renewable Energy Laboratory provided by Department of Energy’s Office of Energy Efficiency and Renewable Energy. Qi Zhang acknowledges funding from National Science Foundation (NSF) (award IIS-2237963).
The opinions, recommendations, findings, and conclusions of this work do not necessarily reflect the views or policies of NIST or the United States Government.

\clearpage
\bibliography{main}

\end{document}